\documentclass[12pt]{article}
\usepackage{amsmath}
\usepackage{bm}
\usepackage{color}

\begin{document}
\begin{center}
\begin{large}
{\bf Kinematic variables in noncommutative phase space and the parameters of noncommutativity}
\end{large}
\end{center}

\centerline {Kh. P. Gnatenko \footnote{E-Mail address: khrystyna.gnatenko@gmail.com}}
\medskip
\centerline {\small \it Ivan Franko National University of Lviv, Department for Theoretical Physics,}
\centerline {\small \it 12 Drahomanov St., Lviv, 79005, Ukraine}

\begin{abstract}
We consider a space with noncommutativity of coordinates and noncommutativity of momenta. It is shown that coordinates in noncommutative phase space depend on mass therefore they can not be considered as kinematic variables. Also, noncommutative momenta are not proportional to a mass as it has to be. We find conditions on the parameters of noncommutativity on which these problems are solved. It is important that on the same conditions the weak equivalence principle is not violated, the properties of kinetic energy are recovered, and the motion of the center-of-mass of composite system and relative motion  are independent in noncommutative phase space.

\end{abstract}

\section{Introduction}	
In recent years many physicists have been worked on the problems in the framework of noncommutative quantum and classical mechanics. Such an interest is motivated by the development of String Theory and Quantum Gravity (see, for example, \cite{Witten,Doplicher}). Idea of noncommutativity was suggested by Heisenberg. The first paper on this subject was written by Snyder in 1947 \cite{Snyder}.

 Different problems were studied in a space with noncommutative algebra, among them are harmonic oscillator \cite{Hatzinikitas,Kijanka,Jing,Smailagic,Smailagic1,Djemai,Giri,Geloun,Nath}, many-particle systems \cite{Djemai,Ho,Daszkiewicz,Gnatenko1,Gnatenko2,Daszkiewicz2,Gnatenko9}, hydrogen atom \cite{Ho,Djemai,Chaichian,Chaichian1,Bertolami,Chair,Stern,Zaim2,Adorno,Khodja,Alavi,Gnatenko6,GnatenkoKr,GnatenkoConf,Balachandran}, Landau problem  \cite{Gamboa1,Horvathy,Dayi,Daszkiewicz1}, gravitational quantum well \cite{Bertolami1,Bastos}, classical systems with various potentials \cite{Gamboa,Romero,Mirza,Djemai1,Gnatenko5,Gnatenko10}, quantum fields  \cite{Balachandran10,Balachandran11} and many others.

Four dimensional noncommutative phase space (2D configurational space and 2D momentum space) can be realized with the help of the following commutation relations for coordinates and momenta
  \begin{eqnarray}
[X_{1},X_{2}]=i\hbar\theta,\label{alI0}\\{}
[X_{i},P_{j}]=i\hbar\delta_{ij},\\{}
[P_{1},P_{2}]=i\hbar\eta,\label{alI1}
\end{eqnarray}
here $\theta$, $\eta$ are constants, called parameters of coordinate and momentum noncommutativity,  $i,j=(1,2)$.

It is worth noting that noncommutativity causes fundamental problems among them  for example are the problem of  rotational symmetry breaking  \cite{Chaichian,Balachandran1,Gnatenko11}, violation of the equivalence principle  \cite{Bastos1,Bertolami2,Gnatenko1,Gnatenko10}, violation of the properties of the kinetic energy  \cite{Gnatenko1,Gnatenko10} and others.
In our previous  paper \cite{Gnatenko10} we studied the problem of violation of the equivalence principle, the problem of violation of the properties of the kinetic energy, the problem of dependence of motion of the center-of-mass of composite system on the relative motion in noncommutative phase space.  We have shown that all these problems are solved if only two conditions on the parameters of noncommutativity are satisfied
 \begin{eqnarray}
\theta_a m_a=\gamma=const,\label{ondN}\\
\frac{\eta_a}{m_a}=\alpha=const,\label{ondN2}
\end{eqnarray}
here $\theta_a$, $\eta_a$ are parameters of noncommutativity which correspond to a particle of mass $m_a$, $\alpha$ and $\gamma$ are constants which are the same for particles with different masses.

In the present paper we show that in the case when  conditions  (\ref{ondN}), (\ref{ondN2}) hold noncommutative coordinates do not depend on mass and can be considered as  kinematic variables, also noncommutative momenta are proportional to a mass as it has to be. Therefore the list of problems which can be solved considering conditions (\ref{ondN}), (\ref{ondN2}) is extended.

 The article is organized as follows. In Section 2 we consider representation for coordinates and momenta  in noncommutative phase space. We present conditions on the parameters of noncommutativity on which noncommutative coordinates can be treated as kinematic variables and noncommutative momenta are proportional to mass as it has to be.  In Section 3 we consider coordinates and momenta of the center-of-mass of composite system in noncommutative phase space and analyze  representations for them. Conclusions are presented in Section 4.

\section{Representation for noncommutative  coordinates and noncommutative momenta and parameters of noncommutativity}

The coordinates $X_i$ and the momenta $P_i$ which satisfy noncommutative algebra (\ref{alI0})-(\ref{alI1})
 can be represented by the coordinates $x_i$ and momenta $p_i$ satisfying the ordinary commutation relations
\begin{eqnarray}
[x_{i},x_{j}]=0,\\{}
[p_{i},p_{j}]=0,\\{}
[x_{i},p_{j}]=i\hbar\delta_{ij},{}
\end{eqnarray}
here $i,j=(1,2)$.
The representation has the following form (see  \cite{Bertolami3})
\begin{eqnarray}
X_{1}=\varepsilon(x_{1}-\frac{1}{2}\theta'{p}_{2}),\label{for01010}\\
X_{2}=\varepsilon(x_{2}+\frac{1}{2}\theta'{p}_{1}),\\
P_{1}=\varepsilon(p_{1}+\frac{1}{2}\eta'{x}_{2}),\\
P_{2}=\varepsilon(p_{2}-\frac{1}{2}\eta'{x}_{1}),\label{for010122}
\end{eqnarray}
where
\begin{eqnarray}
\varepsilon=\frac{1}{\sqrt{1+\frac{\theta'\eta'}{4}}},\\
\end{eqnarray}
and $\theta'$, $\eta'$ are constants. Parameters $\theta'$, $\eta'$  are related with $\theta$, $\eta$  as follows
\begin{eqnarray}
\theta=\frac{\theta'}{1+\frac{\theta'\eta'}{4}},\label{form998}\\
\eta=\frac{\eta'}{1+\frac{\theta'\eta'}{4}}.\label{form9988}
\end{eqnarray}
From (\ref{form998})-(\ref{form9988}) we obtain
\begin{eqnarray}
\theta'=\frac{2}{\eta}(1\pm\sqrt{1-\theta\eta}),  \label{991}\\
\eta'=\frac{2}{\theta}(1\pm\sqrt{1-\theta\eta}).\label{9991}
\end{eqnarray}
Taking into account (\ref{for01010})-(\ref{for010122}), (\ref{991}), (\ref{9991}),  we can write
\begin{eqnarray}
X_{1}=\sqrt{\frac{\theta\eta}{2(1\pm\sqrt{1-\theta\eta})}}\left(x_{1}-\frac{1}{\eta}\left(1\pm\sqrt{1-\theta\eta}\right){p}_{2}\right),\label{rep0}\\
X_{2}=\sqrt{\frac{\theta\eta}{2(1\pm\sqrt{1-\theta\eta})}}\left(x_{2}+\frac{1}{\eta}\left(1\pm\sqrt{1-\theta\eta}\right){p}_{1}\right),\label{rep01}\\
P_{1}=\sqrt{\frac{\theta\eta}{2(1\pm\sqrt{1-\theta\eta})}}\left(p_{1}+\frac{1}{\theta}\left(1\pm\sqrt{1-\theta\eta}\right){x}_{2}\right),\\
P_{2}=\sqrt{\frac{\theta\eta}{2(1\pm\sqrt{1-\theta\eta})}}\left(p_{2}-\frac{1}{\theta}\left(1\pm\sqrt{1-\theta\eta}\right){x}_{1}\right).\label{rep1}
\end{eqnarray}
So, we have two representations (corresponding to $"+"$ or $"-"$ in (\ref{rep0})-(\ref{rep1})) for the coordinates and the momenta which satisfy commutation relations (\ref{alI0})-(\ref{alI1}).
The representations are connected by the following canonical transformation
\begin{eqnarray}
X^{(-)}_{1}=-\sqrt{\frac{\theta}{\eta}}P^{(+)}_2,\label{op}\\
X^{(-)}_{2}=\sqrt{\frac{\theta}{\eta}}P^{(+)}_1,\\
P^{(-)}_{1}=\sqrt{\frac{\eta}{\theta}}X^{(+)}_2,\\
P^{(-)}_{2}=-\sqrt{\frac{\eta}{\theta}}X^{(+)}_1,\label{op1}
\end{eqnarray}
here we use notations $X^{(+)}_{i}$, $P^{(+)}_{i}$ for coordinates and momenta in the case when the sign "+" is chosen in formulas (\ref{rep0})-(\ref{rep1}) and  $X^{(-)}_{i}$, $P^{(-)}_{i}$ for coordinates and momenta in the case of chouse of the sign "-".

In the limits $\theta\rightarrow0$, $\eta\rightarrow0$ from (\ref{rep0})-(\ref{rep1}) we have
\begin{eqnarray}
X^{(-)}_{i}=x_i,\label{X}\\
P^{(-)}_{i}=p_i.\label{P}
 \end{eqnarray}
 So, one has coordinates $x_i$ and momenta $p_i$ which satisfy the ordinary commutation relations as it has to be. In the case when the sign "+" in (\ref{rep0})-(\ref{rep1}) is chosen in the limits $\theta\rightarrow0$, $\eta\rightarrow0$  we have
 \begin{eqnarray}
X^{(+)}_{1}=-\sqrt{\frac{\theta}{\eta}}p_2,\label{can1}\\
X^{(+)}_{2}=\sqrt{\frac{\theta}{\eta}}p_1,\\
P^{(+)}_{1}=\sqrt{\frac{\eta}{\theta}}x_2,\\
P^{(+)}_{2}=-\sqrt{\frac{\eta}{\theta}}x_1,\label{can2}
\end{eqnarray}
 Taking into account (\ref{X}), (\ref{P}), canonical transformation (\ref{can1})-(\ref{can2}) is in agreement with (\ref{op})-(\ref{op1}).

It is important to note that according to  (\ref{rep0})-(\ref{rep01}) the coordinates depend on the momenta $p_i$ and therefore depend on mass. So, the coordinates in noncommutative phase space can not be treated as a kinematic variables.

In general case coordinates
and momenta of different particles may satisfy noncommutative algebra with different parameters of noncommutativity. Let us consider  conditions (\ref{ondN}), (\ref{ondN2}) which relate parameters of noncommutativity which correspond to a particle with its mass.
Let us stress that in the case when conditions  (\ref{ondN}), (\ref{ondN2}) are satisfied we have
\begin{eqnarray}
X_{1}=\sqrt{\frac{\alpha\gamma}{2(1\pm\sqrt{1-\alpha\gamma})}}\left(x_{1}-\frac{1}{\alpha}\left(1\pm\sqrt{1-\alpha\gamma}\right)\frac{{p}_{2}}{m}\right),\label{ep0}\\
X_{2}=\sqrt{\frac{\alpha\gamma}{2(1\pm\sqrt{1-\alpha\gamma})}}\left(x_{2}+\frac{1}{\alpha}\left(1\pm\sqrt{1-\alpha\gamma}\right)\frac{{p}_{1}}{m}\right),\label{ep01}\\
P_{1}=\sqrt{\frac{\alpha\gamma}{2(1\pm\sqrt{1-\alpha\gamma})}}\left(p_{1}+\frac{m}{\gamma}\left(1\pm\sqrt{1-\alpha\gamma}\right){x}_{2}\right),\label{ep111}\\
P_{2}=\sqrt{\frac{\alpha\gamma}{2(1\pm\sqrt{1-\alpha\gamma})}}\left(p_{2}-\frac{m}{\gamma}\left(1\pm\sqrt{1-\alpha\gamma}\right){x}_{1}\right).\label{ep1}
\end{eqnarray}
So, if parameters of noncommutativity, corresponding to a particle, are determined by its mass $m$ as
 \begin{eqnarray}
\theta=\frac{\gamma}{m},\label{ndN}\\
\eta=\alpha m,\label{ndN2}
\end{eqnarray}
the coordinates $X_{i}$ do not depend on the mass of a particle and can be considered as a kinematic variables in noncommutative phase space. Note also that in the case when relations (\ref{ondN}), (\ref{ondN2}) hold we have that momenta (\ref{ep111}), (\ref{ep1}) are proportional to mass as it has to be.

At the end of this section let us consider also the case when the constants in (\ref{for01010})-(\ref{for010122}) are chosen as follows $\varepsilon=1$, $\eta'=\eta$, $\theta'=\theta$  \cite{Bertolami3}. As a result, we can write the following representation
\begin{eqnarray}
X_{1}=x_{1}-\frac{1}{2}\theta{p}_{2},\label{fom01010}\\
X_{2}=x_{2}+\frac{1}{2}\theta{p}_{1},\\
P_{1}=p_{1}+\frac{1}{2}\eta{x}_{2},\\
P_{2}=p_{2}-\frac{1}{2}\eta{x}_{1},\label{fom010122}
\end{eqnarray}
 Commutation relations for coordinates and momenta represented as (\ref{fom01010})-(\ref{fom010122}) reproduce (\ref{alI0}), (\ref{alI1}),
but the commutator of $X_i$ and $P_i$ reads
  \begin{eqnarray}
[X_{i},P_{j}]=i\hbar_{eff}\delta_{ij},
\end{eqnarray}
here
\begin{eqnarray}
h_{eff}=\hbar\left(1+\frac{\theta\eta}{4}\right).
\end{eqnarray}
is called effective Planck constant \cite{Bertolami3}.
Note that if conditions (\ref{ondN}), (\ref{ondN2}) are satisfied the coordinates and momenta can be written as
\begin{eqnarray}
X_{1}=x_{1}-\frac{1}{2}\gamma\frac{{p}_{2}}{m},\label{form01010}\\
X_{2}=x_{2}+\frac{1}{2}\gamma\frac{{p}_{1}}{m},\\
P_{1}=p_{1}+\frac{1}{2}\alpha m{x}_{2},\\
P_{2}=p_{2}-\frac{1}{2}\alpha m{x}_{1},\label{form010122}
\end{eqnarray}
So, in this case coordinates do not depend on the mass and can be treated as kinematic variables. Also, momenta are proportional to the mass, as it has to be. It is worth also mention that in the case when parameters of noncommutativity are determined by mass (\ref{ondN}), (\ref{ondN2}) the effective Planck constant reads
\begin{eqnarray}
h_{eff}=\hbar\left(1+\frac{\alpha\gamma}{4}\right).
\end{eqnarray}
 and is the same for different particles.

 In the next section we will show that conditions are also important in studying of composite system in noncommutative phase space.

\section{Coordinates and momenta of the center-of-mass of composite system}

In general case coordinates and momenta of different particles may feel noncommutativity with different parameters
\begin{eqnarray}
[X_{1}^{(a)},X_{2}^{(b)}]=i\hbar\delta^{ab}\theta_{a},\label{al0}\\{}
[X_{i}^{(a)},P_{j}^{(b)}]=i\hbar\delta^{ab}\delta_{ij},\\{}
[P_{1}^{(a)},P_{2}^{(b)}]=i\hbar\delta^{ab}\eta_{a},\label{al1}
\end{eqnarray}
here indices $a$, $b$ label the particles, $i=(1,2)$, $j=(1,2)$, $\theta_{a}$, $\eta_{a}$ are parameters of noncommutativity, which correspond to a particle of mass $m_a$. Therefore there is a problem of describing
the motion of the center-of-mass of the composite system in noncommutative phase space. This problem was studied in our previous paper \cite{Gnatenko10}. We showed that coordinates and momenta of the center-of-mass of composite system
\begin{eqnarray}
\tilde{{\bf X}}=\frac{\sum_{a}m_{a}{\bf X}^{(a)}}{\sum_{a}m_{a}},\label{xcm}\\
\tilde{{\bf P}}=\sum_{a}{\bf P}^{(a)},\label{pcm}
\end{eqnarray}
satisfy noncommutative algebra with effective parameters of noncommutativity $\tilde{\theta}$, $\tilde{\eta}$. Taking into account (\ref{al0})-(\ref{al1})
and (\ref{xcm})-(\ref{pcm}) one has
\begin{eqnarray}
[\tilde{X}_1,\tilde{X}_2]=i\hbar\tilde{\theta},\label{07}\\{}
[\tilde{P}_1,\tilde{P}_2]=i\hbar\tilde{\eta},\\{}
[\tilde{X}_i,\tilde{P}_j]=i\hbar\delta_{ij},\label{08}{}
\end{eqnarray}
with
\begin{eqnarray}
\tilde{\theta}=\frac{\sum_{a}m_{a}^{2}\theta_{a}}{(\sum_{b}m_{b})^{2}},\label{eff}\\
\tilde{\eta}=\sum_{a}\eta_a.\label{eff2}
\end{eqnarray}

Analogically to (\ref{rep0})-(\ref{rep1}),  the coordinates and momenta of the center-of-mass which satisfy (\ref{07})-(\ref{08}) can be represented as
\begin{eqnarray}
\tilde{X}_{1}=\sqrt{\frac{\tilde{\theta}\tilde{\eta}}{2(1\pm\sqrt{1-\tilde{\theta}\tilde{\eta}})}}\left(\tilde{x}_{1}-\frac{1}{\tilde{\eta}}\left(1\pm\sqrt{1-\tilde{\theta}\tilde{\eta}}\right)\tilde{p}_{2}\right),\label{repc0}\\
\tilde{X}_{2}=\sqrt{\frac{\tilde{\theta}\tilde{\eta}}{2(1\pm\sqrt{1-\tilde{\theta}\tilde{\eta}})}}\left(\tilde{x}_{2}+\frac{1}{\tilde{\eta}}\left(1\pm\sqrt{1-\tilde{\theta}\tilde{\eta}}\right)\tilde{p}_{1}\right),\label{repc01}\\
\tilde{P}_{1}=\sqrt{\frac{\tilde{\theta}\tilde{\eta}}{2(1\pm\sqrt{1-\tilde{\theta}\tilde{\eta}})}}\left(\tilde{p}_{1}+\frac{1}{\tilde{\theta}}\left(1\pm\sqrt{1-\tilde{\theta}\tilde{\eta}}\right)\tilde{x}_{2}\right),\label{repc02}\\
\tilde{P}_{2}=\sqrt{\frac{\tilde{\theta}\tilde{\eta}}{2(1\pm\sqrt{1-\tilde{\theta}\tilde{\eta}})}}\left(\tilde{p}_{2}-\frac{1}{\tilde{\theta}}\left(1\pm\sqrt{1-\tilde{\theta}\tilde{\eta}}\right)\tilde{x}_{1}\right).\label{repc1}
\end{eqnarray}
here
\begin{eqnarray}
\tilde{x}_i=\frac{\sum_{a}m_{a}{x}_i^{(a)}}{\sum_{a}m_{a}},\label{xcm1}\\
\tilde{p}_i=\sum_{a}{p}_i^{(a)},\label{pcm1}
\end{eqnarray}
are coordinates and momenta of the center-of-mass
which satisfy
\begin{eqnarray}
[\tilde{x}_{i},\tilde{x}_{j}]=0,\\{}
[\tilde{p}_{i},\tilde{p}_{j}]=0,\\{}
[\tilde{x}_{i},\tilde{p}_{j}]=i\hbar\delta_{ij}.{}
\end{eqnarray}

 On the other hand, representation for coordinates $X_{1}^{(a)}$ and momenta $P_{1}^{(a)}$ of a particle with parameters $\theta_a$ and $\eta_a$ is given by (\ref{rep0})-(\ref{rep1}).
Substituting  (\ref{rep0})-(\ref{rep1}) into  (\ref{xcm}), (\ref{pcm}), we have
\begin{eqnarray}
\tilde{X}_{1}=\frac{1}{M}\sum_a m_a\sqrt{\frac{\theta_a\eta_a}{2(1\pm\sqrt{1-\theta_a\eta_a})}}\left(x^{(a)}_{1}-\frac{1}{\eta_a}\left(1\pm\sqrt{1-\theta_a\eta_a}\right){p}^{(a)}_{2}\right),\label{reps0}\\
\tilde{X}_{2}=\frac{1}{M}\sum_a m_a \sqrt{\frac{\theta_a\eta_a}{2(1\pm\sqrt{1-\theta_a\eta_a})}}\left(x^{(a)}_{2}+\frac{1}{\eta_a}\left(1\pm\sqrt{1-\theta_a\eta_a}\right){p}^{(a)}_{1}\right),\label{reps01}\\
\tilde{P}_{1}=\sum_a \sqrt{\frac{\theta_a\eta_a}{2(1\pm\sqrt{1-\theta_a\eta_a})}}\left(p^{(a)}_{1}+\frac{1}{\theta_a}\left(1\pm\sqrt{1-\theta_a\eta_a}\right){x}^{(a)}_{2}\right),\label{reps02}\\
\tilde{P}_{2}=\sum_a \sqrt{\frac{\theta_a\eta_a}{2(1\pm\sqrt{1-\theta_a\eta_a})}}\left(p^{(a)}_{2}-\frac{1}{\theta_a}\left(1\pm\sqrt{1-\theta_a\eta_a}\right){x}^{(a)}_{1}\right).\label{reps1}
\end{eqnarray}
here $M$ is the total mass of the system $M=\sum_{a}m_{a}$.
Note that
 representations (\ref{repc0})-(\ref{repc1}) and (\ref{reps0})-(\ref{reps1}) are not the same.

It is interesting to mentioning that if conditions (\ref{ondN}), (\ref{ondN2}) are satisfied expressions  (\ref{reps0})-(\ref{reps1}) reproduce (\ref{repc0})-(\ref{repc1}). We have
 \begin{eqnarray}
\tilde{X}_{1}=\sqrt{\frac{\alpha\gamma}{2(1\pm\sqrt{1-\alpha\gamma})}}\left(\tilde{x}_{1}-\frac{1}{\alpha}\left(1\pm\sqrt{1-\alpha\gamma}\right)\frac{\tilde{p}_{2}}{M}\right),\label{r0}\\
\tilde{X}_{2}=\sqrt{\frac{\alpha\gamma}{2(1\pm\sqrt{1-\alpha\gamma})}}\left(\tilde{x}_{2}+\frac{1}{\alpha}\left(1\pm\sqrt{1-\alpha\gamma}\right)\frac{\tilde{p}_{1}}{M}\right),\label{r01}\\
\tilde{P}_{1}=\sqrt{\frac{\alpha\gamma}{2(1\pm\sqrt{1-\alpha\gamma})}}\left(\tilde{p}_{1}+\frac{M}{\gamma}\left(1\pm\sqrt{1-\alpha\gamma}\right)\tilde{x}_{2}\right),\label{r02}\\
\tilde{P}_{2}=\sqrt{\frac{\alpha\gamma}{2(1\pm\sqrt{1-\alpha\gamma})}}\left(\tilde{p}_{2}-\frac{M}{\gamma}\left(1\pm\sqrt{1-\alpha\gamma}\right)\tilde{x}_{1}\right).\label{r1}
\end{eqnarray}
where $\tilde{x}_{i}$, $\tilde{p}_{i}$ are defined by (\ref{xcm1}), (\ref{pcm1}). Here we use the following relations
 \begin{eqnarray}
\tilde{\theta}=\frac{\gamma}{M},\label{N}\\
\tilde{\eta}=\alpha M,\label{N2}
\end{eqnarray}
which are obtained from (\ref{ondN}), (\ref{ondN2}), (\ref{eff}), (\ref{eff2}).

It is important to note that in the contrast to (\ref{repc02})-(\ref{repc1}), (\ref{reps02})-(\ref{reps1})
 momenta (\ref{r02})-(\ref{r1}) are proportional to the total mass $M$. So, conditions (\ref{ondN}), (\ref{ondN2}) give the possibility to recover proportionality of total momenta to $M$.

The same conclusion can be done in the case of representation
 (\ref{form01010})-(\ref{form010122}).
In analogy to (\ref{repc0})-(\ref{repc1}), (\ref{reps0})-(\ref{reps1}) the coordinates and momenta of the center-of-mass can be represented as
\begin{eqnarray}
\tilde{X}_{1}=\tilde{x}_{1}-\frac{1}{2}\tilde{\theta}\tilde{p}_{2},\label{f09}\\
\tilde{X}_{2}=\tilde{x}_{2}+\frac{1}{2}\tilde{\theta}\tilde{p}_{1},\\
\tilde{P}_{1}=\tilde{p}_{1}+\frac{1}{2}\tilde{\eta}\tilde{x}_{2},\label{f0009}\\
\tilde{P}_{2}=\tilde{p}_{2}-\frac{1}{2}\tilde{\eta}\tilde{x}_{1},\label{f009}
\end{eqnarray}
and
\begin{eqnarray}
\tilde{X}_{1}=\sum_a \frac{m_a}{M} \left(x^{(a)}_{1}-\frac{1}{2}\theta_a{p}^{(a)}_{2}\right),\label{f10}\\
\tilde{X}_{2}=\sum_a \frac{m_a}{M} \left(x^{(a)}_{2}+\frac{1}{2}\theta_a{p}^{(a)}_{1}\right),\\
\tilde{P}_{1}=\sum_a \left(p^{(a)}_{1}+\frac{1}{2}\eta_a{x}^{(a)}_{2}\right),\label{f0010}\\
\tilde{P}_{2}=\sum_a \left(p^{(a)}_{2}-\frac{1}{2}\eta_a{x}^{(a)}_{1}\right),\label{f010}
\end{eqnarray}
 Representations (\ref{f09})-(\ref{f009}) and (\ref{f10})-(\ref{f010}) are not the same. Note that when conditions (\ref{ondN}), (\ref{ondN2}) hold from (\ref{f09})-(\ref{f009}) and (\ref{f10})-(\ref{f010}) we obtain the following expressions for noncommutative coordinates and nocommutative momenta
\begin{eqnarray}
\tilde{X}_{1}=\tilde{x}_{1}-\frac{1}{2}\gamma\frac{\tilde{p}_{2}}{M},\label{form0}\\
\tilde{X}_{2}=\tilde{x}_{2}+\frac{1}{2}\gamma\frac{\tilde{p}_{1}}{M},\\
\tilde{P}_{1}=\tilde{p}_{1}+\frac{1}{2}\alpha M\tilde{x}_{2},\label{form00}\\
\tilde{P}_{2}=\tilde{p}_{2}-\frac{1}{2}\alpha M\tilde{x}_{1},\label{form000}
\end{eqnarray}

It is worth mentioning that in the contrast to (\ref{f0009})-(\ref{f009}) and (\ref{f0010})-(\ref{f010}),  one has  proportionality of the total momenta $\tilde{P}_{i}$ (\ref{form00}), (\ref{form000})  to the total mass $M$.

\section{Conclusions}

In the paper we have considered a space with noncommutativity of coordinates and noncommutativity of momenta.
It is shown that coordinates in noncommutative phase space can not be considered as kinematic variables because of they dependence on the mass.
We have studied a general case when different particles satisfy noncommutative algebra with different parameters of noncommutativity.   We have shown that if parameters of noncommutativity which correspond to a particle are determined by its mass as  (\ref{ondN}), (\ref{ondN2}) the noncommutative coordinates can be treated as kinematic variables.
Moreover, we have shown that noncommutative momenta are proportional to  mass in the case when relations  (\ref{ondN}), (\ref{ondN2}) are satisfied.

In addition coordinates and momenta of the center-of-mass of composite system have been considered in noncommutative phase space.  We have concluded that if conditions (\ref{ondN}), (\ref{ondN2}) hold the total momenta of the system are proportional to its total mass as it has to be. Also, it has been shown that representation  for  coordinates of the center-of-mass and total momenta  (\ref{reps0})-(\ref{reps1}) written on the basis of they definitions (\ref{xcm}), (\ref{pcm}) and representation (\ref{repc0})-(\ref{repc1}) obtained from noncommutative algebra (\ref{07})-(\ref{08}) reproduce each other in the case when relations (\ref{ondN}), (\ref{ondN2}) are satisfied.

So, in the case when parameters of noncommutativity which correspond to a particle are determined by its mass a list of important results can be obtained in noncommutative phase space. In our previous paper \cite{Gnatenko10} we showed that if conditions  (\ref{ondN}), (\ref{ondN2}) hold  the weak equivalence principle is recovered; the properties of the kinetic energy are preserved; the motion of the center-of-mass of composite system and relative motion are independent in noncommutative phase space.
I addition  in this paper we have shown that in the case when conditions (\ref{ondN}), (\ref{ondN2}) are preserved the noncommutative coordinates do not depend on mass and can be treated as kinematic variables in noncommutative phase space, noncommutative momenta are proportional to the mass as it has to be.
So, the importance of proposed conditions (\ref{ondN}), (\ref{ondN2}) is stressed by the number of problems which can be solved in noncommutative phase space.

We wound like also to note that  similar to (\ref{ondN}), (\ref{ondN2}) condition  which relates parameter of deformation $\beta$ with mass  (namely $m\sqrt{\beta}=\gamma$, $\gamma$ is constant) is important in solving problems in deformed space with minimal length $[\hat{X},\hat{P}]=i\hbar(1+\beta\hat{P}^{2})$. Among them are violation of the equivalence principle, violation of properties of kinetic energy, dependence of Galilean an Lorentz transformations on mass  \cite{Tkachuk,Quesne,Tkachuk1}. So, idea to connect parameters of deformations (noncommutative parameters) with mass of a particle is important in solving problems in quantized spaces.

\section*{Acknowledgments}
The author thanks Professor V. M. Tkachuk for his advices and great support during research studies. This work was supported in part by the European Commission under the project STREVCOMS PIRSES-2013-612669 and the project FF-30F (No. 0116U001539) from the
Ministry of Education and Science of Ukraine.

\end{document}